\documentclass[%
 reprint,
 amsmath,amssymb,
 aps,
]{revtex4-2}

\usepackage{graphicx}
\usepackage{dcolumn}
\usepackage{bm}
\usepackage[normalem]{ulem}
\usepackage{xcolor}

\begin{document}

\title{L\'evy flights of photons with infinite mean free path}

\author{Michelle O. Ara\'ujo$^{1,*}$, Thierry Passerat de Silans$^{1,2}$, and Robin Kaiser$^1$}

\affiliation{$^{1}$Universit\'e C\^ote d'Azur, CNRS, Institut de Physique de Nice, F-06560 Valbonne France\\
$^{*}$Present address: Departamento de F\'isica, Universidade Federal de Pernambuco, 50670-901, Recife - PE, Brazil\\
$^2$Departamento de F\'isica/CCEN, Universidade Federal da Para\'iba, Caixa Postal 5008, 58051-900, Jo\~ao Pessoa, PB, Brazil}

\date{\today}
\begin{abstract} 
Multiple scattering of light by resonant vapor is characterized by L\'evy-type superdiffusion with a single-step size distribution $p(x)\propto 1/x^{1+\alpha}$. We investigate L\'evy flight of light in a hot rubidium vapor collisional-broadened by 50 torr of He gas. The frequent collisions produce Lorentzian absorptive and emissive profiles with $\alpha<1$ and a corresponding divergent mean step size. We extract the L\'evy parameter $\alpha\approx0.5$ in a multiple scattering regime from radial profile of the transmission and from violation of the Ohm's law. The measured radial transmission profile and the total diffusive transmission curves are well reproduced by numerical simulations for Lorentzian line shapes.
\end{abstract}

\maketitle

The random walk of particles can often be described within the central limit theorem, which characterizes diffusion phenomena and implies that the mean squared displacement performed by a particle increases linearly in time. However, many physical systems exhibit superdiffusion, in which the mean squared displacement grows faster than a linear function of time \cite{Dubkov2008}. A particular mechanism for superdiffusion is L\'evy flights, where rare long jumps dominate the dynamics of the random walk. 
The occurrence of L\'evy flights is not rare \cite{Frisch1995, Shlesinger1999, Viswanathan2008, Miron2020} and they are encountered in a variety of systems, ranging from human travel \cite{Brockmann2006,Gonzales2008,Reynolds2018} spread of diseases \cite{Hilhorst1999, Torcini2006, DeVerDye2020,Havlin2020}, trajectories of animals \cite{Viswanathan1996,Edwards2007,Uzeda2019}, turbulence \cite{Shlesinger1987} and financial market \cite{Podobnik2011}. 

For jump size distributions $p(x)$ decaying asymptotically with $p(x)\propto x^{-(1+\alpha)}$, it is straightforward to show that the second moment of the jump size $\left\langle x^2 \right\rangle$ is finite for $\alpha > 2$, and the central limit theorem will then apply. For $\alpha < 2$, $\left\langle x^2 \right\rangle$ becomes infinite, and the random walk can no longer be described by a diffusion equation. This is the regime of L\'evy flights. An even more extreme regime corresonds to $\alpha < 1$, where even the average jump size (or scattering mean free path) diverges. 

Long jumps have been recognized as an important mechanism for understanding light transport in scattering media almost 100 years ago \cite{Kenty1932} and are at the basis of many radiative transfert codes used in astrophysics \cite{Hummer1968}.
Modern experimental development allowed for the investigation of L\'evy flights of light in controllable and tunable systems. For instance, L\'evy flights were investigated in engineered media \cite{Barthelemy2008, Bertolotti2010} and from light diffusion in atomic vapors \cite{Mercadier2009,Mercadier2013,Baudouin2014}. The control of the optical system allows the investigation of how the light transport is impacted by, e.g., quenched \cite{Barthelemy2010,Svensson2013,Svensson2014} and annealed \cite{Baudouin2014} disorder, correlations induced by inelastic scattering \cite{Pereira2007, Carvalho2015}, fractal dimension of the random walk \cite{Savo2014} and effects of system size \cite{Savo2014}.

Light transport in resonant vapors is known to depend on the absorption profile and the frequency redistribution between the scattering processes. 
The asymptotic decay $p(x)\approx x^{-(1+\alpha)}$ is expected to scale as $\alpha=1$ for Doppler broadened vapors and as $\alpha=0.5$ for Lorentz ones \cite{Pereira2004}. The experimental measurement of the L\'evy exponent $\alpha$ for a Doppler vapor was done both directly, by means of the measurement of the jump size distribution \cite{Mercadier2009,Mercadier2013} and indirectly, analyzing transmission signatures in the multiple scattering regime \cite{Baudouin2014}. Indeed the radial profile $T(r)$ scales with $T(r)\propto r^{-(3+\alpha)}$ for radial distances $r$ larger than the sample thickness. Moreover, the total scattered light $T_\mathrm{diff}$ scales with the sample opacity $O$ as $ T_\mathrm{diff} \propto O^{-\alpha/2}$ \cite{Groth2012,Baudouin2014}.

In this work, we report on the realization of L\'evy flights of light in hot atomic vapors with a L\'evy exponent $\alpha \approx 0.5$ corresponding to a regime of infinite scattering mean free path. The modification of the L\'evy exponent is obtained by admixing a buffer gas of He atoms into a Rubidium vapor. Following \cite{Barthelemy2008, Baudouin2014} we extract the L\'evy exponent  $\alpha$ by analysing the radial profile of the transmitted light and the total diffuse transmission for a multiple scattering regime. 


\begin{figure}[hbtp]
\centering
\includegraphics{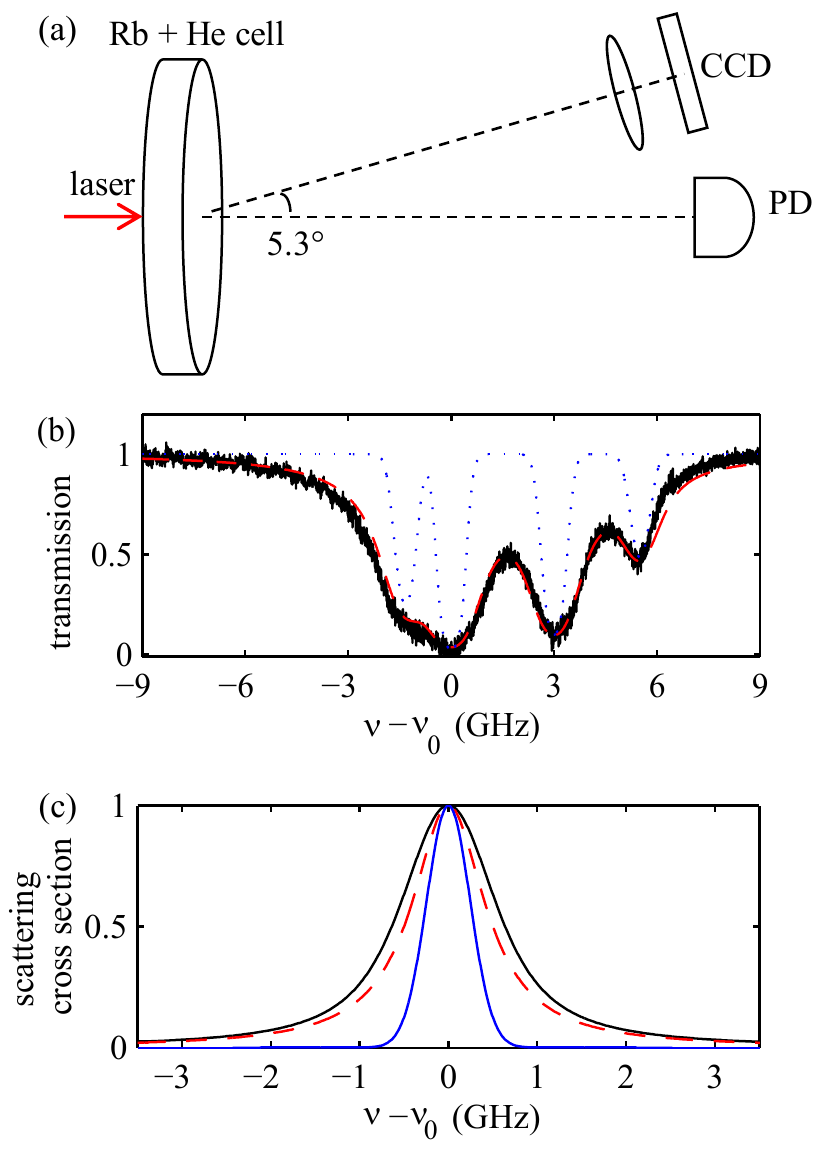}
\caption{
(a) Experimental setup. A resonant laser beam excites a disk-shaped cell filled with Rb vapor and 50 torr of He gas. A photodiode (PD) records the coherent transmission and a CCD camera collects the scattered light. 
(b) Experimental coherent transmission spectrum (black line) and theoretical fits using Voigt (red dashed line) or Doppler profiles (blue dotted line) for a temperature $T=106^{\circ}$C. The Voigt-profile fit gives an opacity $O = 11$.
(c) Calculated normalized scattering cross section for a single $^{85}$Rb line using a Voigt profile (black solid line), a Lorentz profile with same collisional broadening (red dashed line) or a Doppler profile for $T=106^{\circ}$C (blue solid line). 
}
\label{figure1}
\end{figure}


In order to evaluate the potential of controlling L\'evy flights in atomic vapors, we perform numerical simulations by describing the photon random walk in the vapor by successive absorption and emission processes. For each scattering event, the emitted frequency is redistributed and is partially correlated to the absorbed frequency. Two scenarios are usually considered for realistic vapors, namely \cite{Pereira2007}: (i) $R_{II}$ for combined Doppler and natural broadening, and (ii) $R_{III}$ for combined Doppler, natural and collisional broadening. For the $R_{II}$ scenario, the absorption-emission process is elastic in the atomic rest frame but the coherence is partially lost in the laboratory frame due to Doppler shifts. For the $R_{III}$ scenario, coherence in the atomic rest frame is lost due to collision with other atoms (e.g., with the buffer gas) and the emission is Lorentzian in the atomic rest frame. Nevertheless, partial frequency coherence is possible in the laboratory rest frame due to velocity selection in the atomic absorption. When collisions are very frequent, as will be the case in our experiment, a $R_{III}$ scenario will be the most appropriate one. Moreover, when collisional broadening dominates Doppler broadening,  complete frequency redistribution occurs for a single scattering event with a Voigt emission profile with large Lorentz wings \cite{Romalis1997,Kielkopf1975}. 

To characterize the vapor absorption properties, we first compute the dimensionless quantity $f(\delta) = \sigma(\delta)/\sigma_0$, where $\sigma(\delta)$ the scattering cross section at detuning $\delta=\nu-\nu_0$ of the laser frequency $\nu$ from the atomic transition frequency $\nu_0$ and  
$\sigma_0=\frac{3\lambda^2}{2\pi}\frac{\Gamma_0}{\Gamma}$ is the scattering cross section of a pinned two-level atom at resonance. Here $\Gamma=\Gamma_0+\Gamma_C$ is the homogeneously broadened linewidth 
where $\Gamma_0/2\pi= 6$ MHz is the natural width of Rb and $\Gamma_C/2\pi\approx 1$ GHz is the collisional width at 50 torr of He \cite{Romalis1997}.	
We take the sum of four two-level profiles corresponding to two ground states per Rb isotope, so  $f(\delta)=\sum_i\beta_if_i(\delta_i)$, where $\delta_i$ is the detuning relative to transition $i$ and $\beta_i$ is a weight coefficient to take into account the isotope concentration and the coupling strengths for each transition. The individual $f_i(\delta_i)$ are Voigt profiles, i.e., the convolution of a Lorentz line, defined by the collisions with buffer gas, and the Maxwell-Boltzmann velocity for the atoms in the vapor,

\begin{equation}
f_i(\delta_i)=\frac{1}{u\sqrt{\pi}}\int dve^{-v^2/u^2}\frac{1}{1+4\left(\delta_i-ku\right)^2/\Gamma^2} \label{Voigt}
\end{equation} 

where $k=\frac{2\pi}{\lambda}$ is the wavenumber, $u=\sqrt{2k_BT/m}$ is the most probable atomic speed, $k_B$ the Boltzmann constant and $m$ the atomic mass. 
The scattering cross section is a Voigt profile with parameter $a=\Gamma/\Gamma_D\approx 3$, where 
$\Gamma_D/2\pi=u/\lambda$ is the Doppler width. 
This allows us to compute a transmission spectrum of the light through an atomic vapor of density $n$ and length $L$, yielding a resonant opacity $O$, defined as $O=n\sigma_0 L$ \cite{Baudouin2014}, by fitting a Beer-Lambert law 
\begin{equation}
T_\mathrm{coh}=e^{-O\,\sigma(\delta)/\sigma_0}. \label{BeerLambert}
\end{equation}
An example of a coherent transmission spectra fitted with Eqs. \ref{BeerLambert} and \ref{Voigt} is shown in Fig. \ref{figure1}b, with  the opacity as the only fitting parameter. We note that a purely Doppler broadened absorption profile (shown in blue in Fig. \ref{figure1}b does not allow a correct description of our experiment in presence of the He buffer gas.

We now turn to the multiple scattering regime. Within the $R_{III}$ ansatz, the step size distribution \cite{Pereira2004,Pereira2007} is given by:
\begin{equation}
p(\delta',x)=\int d\delta' \Phi(\delta,\delta')\phi(\delta)e^{-\phi(\delta)x}, \label{px}
\end{equation}

with $\Phi(\delta,\delta')$ the probability of having an emission at detuning $\delta$ if the incident photon is at detuning $\delta'$.
For complete frequency redistribuiton (CFR) limit (valid after several scattering events), the emission profile is equal to the absorption one and $p(x)=\int\ d\delta\, \phi(\delta)^2e^{-\phi(\delta)x}$, which decays  asymptotically as $p(x)\rightarrow x^{-2}$ for a Doppler profile and as $p(x)\rightarrow x^{-1.5}$ for a Lorentz profile \cite{Pereira2004}.


\begin{figure}[h]
\centering
\includegraphics[scale=1]{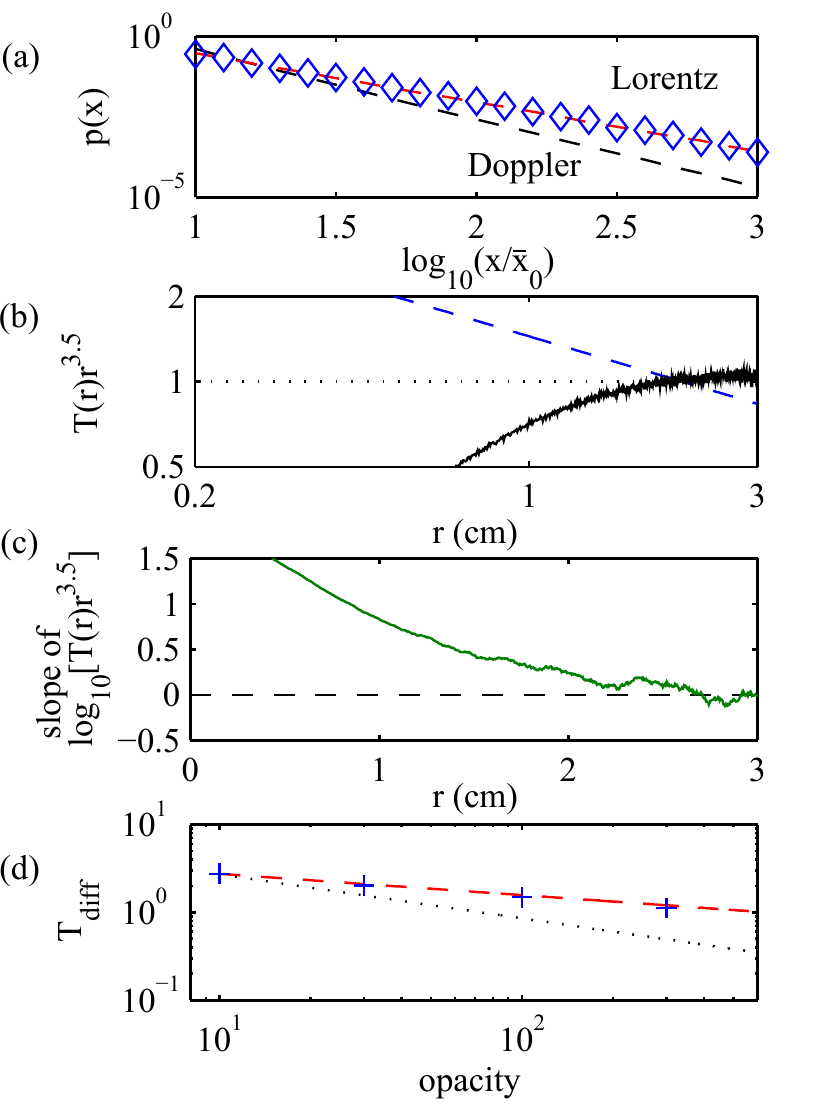}
\caption{
a) Calculated step size distribution $p(x)$ for a $R_{III}$ scenario with Voigt parameter $a=3.0$ (hot Rb cell mixed with 50 torr of He gas) as a function of $log_{10}(x/\bar{x}_0)$, with $\bar{x}_0=1/n\sigma(0)$ the absorption penetration depth at line center. Expected aysmptotic lines are for CFR with Doppler (black) and Lorentz (red) profiles, blue diamonds represent the step size distribution after a single scattering event with incident photon at line center. b) Simulated rescaled radial transmission profile $T(r) r^{3.5}$ for the same cell geometry of the experiment (black line). A rescaled Doppler asymptotic is shown in blue dashed line.
c) Local slope of the rescaled transmission profile. d) Total diffusive transmission: result of Monte Carlo simulation for a Voigt parameter $a=3.0$ (blue crosses) and expected asymptotic behavior (red dashed line). The diffuse transmission curve for Doppler case is shown for comparison (black points).}
\label{figure2}
\end{figure}

To have more insight on the effect of frequency redistribution on the random walk, we have calculated $p(x)$ from Eq. \ref{px} for the $R_{III}$ scenario for a Voigt parameter $a=3$. The calculated $p(x)$ is shown in Fig. \ref{figure2}a together with results for CFR with Doppler and Lorentz profiles. For our Rb cell filled with 50 torr of He gas, $p(x)$ follows the Lorentz limit for almost all opacities, since the frequency is completely redistributed for a Voigt profile with $a\approx3$. 

For the range of opacities used in the experiment, most photons escape the cell after multiple scattering. Still, information about the asymptotic behaviour of the step length distribution subsist on the transmitted light \cite{Baudouin2014}.

\begin{figure}[h]
\centering
\includegraphics[width=1.\linewidth]{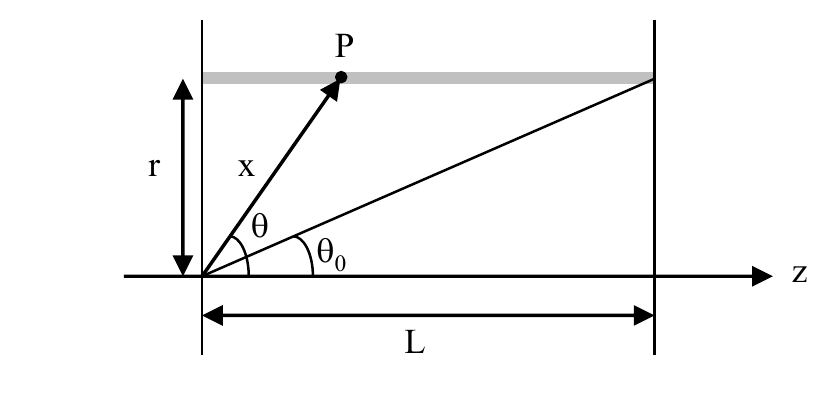}
\caption{Illustrating scheme for theoretical calculation of radial diffusive transmission. The photon is supposed to start from coordinate system origin ($r=0$, $z=0$) and do a large step to point $P$ at radial distance $r$ from longitudinal axis from which it is scattered and detected. We integrate over all possible points $P$ to obtain radial transmission $T(r)$.}
\label{FigEq4}
\end{figure}

The radial intensity distribution of the transmission $T(r)$ for large  $r$ ($r\gg L$) is dominated by single large step $x\gg L$  originated near origin $r\approx 0 $, $z\approx 0$ (see Fig. \ref{FigEq4}) \cite{Baudouin2014}. The photon will be detected after being scattered at a point $P$ with radial distance $r$ to the origin. The probability that the photon reaches point $P$ from origin is $p(x)dx\sin(\theta)d\theta$, the product of the probability of doing a step with length between $x$ and $x+dx$ with the probability of doing an angle between $\theta$ and $\theta+d\theta$.

The number of photons scattered at radial distance between $r$ and $r+dr$ is obtaining integrating over all possible angles with the step length given by $x=r/\sin(\theta)$:
\begin{equation}
T(r)rdr\propto =\frac{1}{r^{1+\alpha}}dr\int_{\theta_0}^{\pi/2}\sin^{1+\alpha}(\theta)d\theta, \label{radial}   
\end{equation}
with $\theta_0$ the minimal angle to reach radial distance $r$ defined as $\cos(\theta_0)=\frac{L}{d}$ and $d=\sqrt{r^2+L^2}$. The integral in right hand side of eq. \ref{radial} has analytical solution resulting in:
\begin{equation}
T(r)rdr\propto \frac{1}{r^{1+\alpha}}dr\left[\frac{L}{d}\: _2F_1\left(0.5,-\alpha/2,1.5,\frac{L^2}{d^2}\right)\right],
\end{equation}
with $_2F_1$ denoting the hypergeometric function.

In the limit $r\gg L$ one has $\frac{L}{d}\rightarrow \frac{L}{r}$ and the hypergeometric function tends to one which allows us to obtain the asymptotic behavior 
\begin{equation}
T(r)\propto r^{-(3+\alpha)}. \label{radialTrans}   
\end{equation}

This asymptotic scaling is confirmed by our Monte-Carlo simulations where we compute the complete transmission profile of the photons through a cell with the geometry used in the experiment and an opacity value of $O=300$. For instance, for this typical opacity value almost $85\%$ of the photons do at least 12 scattering events before being detected (see Supplemental Material \cite{Supp} for probability density function of the number of scattering events undergone by detected photons before escaping the cell). For clarity we show in Fig \ref{figure2}b the rescaled transmission profile $T(r)r^{3.5}$ expected to asymptotically constant as $T(r)\propto r^{-3.5}$ for a Lorentzian vapor ($\alpha=0.5$). A line decaying as $r^{-0.5}$ shows the expected rescaled behavior for a Doppler vapor with $P(r)\propto r^{-4}$\cite{Baudouin2014}. 
In Fig. \ref{figure2}c we show a local slope of the rescaled profile $T(r)r^{3.5}$. The simulation takes into account reflections on the boundaries of the cell and the collection angle of the lens used in experiment. Reflections have the effect of increasing diffusive transmission at large radial distance since those position are reached by photons with large angles relative to the normal direction of the windows but does not affect the asymptotic dependence on $r^{-3-\alpha}$. Simulation results have been shown to be independent on selecting the output photon angle in accordance to experiment detection geometry or taking all photons that escape through the output window (see Supplemental Material \cite{Supp} for simulation tests on the effects of reflection and angle selection).

Information about asymptotic behavoir of the step size distribution $p(x)$ can also be obtained from deviations of the Ohm's law. Indeed, superdiffusion favors the escape of photon relative to normal diffusion and the total transmission decays slower with system size (or opacity $O$) than the Ohm law $T_\mathrm{Ohm}\propto O^{-1}$ in regular diffusive samples. Details of the total diffusive transmission $T_\mathrm{diff}$ with $O$ depend on the input geometry and our experimental set-up corresponds to the so called non-equilibrium initial conditions \cite{Groth2012}, for which the first scattering event occurs at $z\approx 0$. Indeed, in our experiment, the incident photons are resonant with the $F=3$ (or $F=2$ as well) state of $^{85}$Rb for which the absorption length inside the vapor is $l_{a}=L/O\ll L$, so the first scattering event occurs close to the input cell window. Under these conditions, the dependence of $T_\mathrm{diff}$ with opacity $O$ is given by \cite{Groth2012}:
\begin{equation}
T_\mathrm{diff}\propto O^{-\alpha/2}. \label{T_diff}
\end{equation}

The dependency of $T_\mathrm{diff}$ with opacity obtained in the simulations are shown in Fig. \ref{figure2}d together with a lines corresponding to the expected $O^{-0.25}$ behavior for Lorentz vapor and $O^{-0.5}$ expected for Doppler vapor \cite{Baudouin2014}.

In our experiment (Fig. \ref{figure1}(a)), a disk shaped cell of radius $5.0$ cm and internal thickness $L=6.3$ mm is filled with a natural mixture of $^{85}$Rb and $^{87}$Rb isoptopes and $50$ torr of He gas. As already mentioned, this pressure gives a estimated collisional broadening of $\Gamma_C/2\pi\approx 1$ GHz \cite{Romalis1997}, almost three times the Doppler width for the D2 line of Rb at $\lambda=780$ nm and large enough for the scattering cross section to be approximated by a Lorentzian profile. A collimated laser beam from a Ti:Sa laser source of waist $0.88$ mm and negligible linewidth ($\approx 50$kHz) is sent perpendicularly to the cell surface and close to its center to excite the atoms. The power of the laser beam is $2$ $\mu$W, which gives a very low saturation parameter due to the important collisional broadening. Measurements at $1$ $\mu$W yielded similar results, allowing to exclude nonlinear optics effects to occur \cite{Santic2018}. In order to obtain an opaque sample around the line center, we heat the cell between $T=106^{\circ}$C and $T=180^{\circ}$C. Transmission spectra are obtained by scanning the laser frequency across the absorption lines allowing to extract the resonant opacity $O$ using Eq.(\ref{BeerLambert}).
For this temperature range, the resonant opacities vary from $O=11$ to $O=530$ (corresponding to Rb vapor densities from $n=5\times10^{12}$ cm$^{-3}$ to $n=2\times10^{14}$ cm$^{-3}$), with a typical uncertainty of 10\%.



Monitoring the linear absorption spectrum in an auxiliary Rb cell with no buffer gas, 
we tune the frequency to the minimum of transmission of the $F=3$ hyperfine ground state of $^{85}$Rb, taken as the reference frequency $\nu_0$.
A CCD camera from Andor (iXon X3 885, pixel size $8.0$ $\mu$m, exposure time of $1.0$s) placed at an angle of $5.3^{\circ}$ from the transmitted laser beam records images of the scattered light from the output facet (Fig. \ref{figure1}a). A lens of focal distance $50$ mm is placed between the cell and the camera to provide a magnification of $9.4$ for the images.

We then extract the radial profile $T(r)$ of the scattered light at the output facet by performing an angular average of the images around the center of maximum intensity. An image without the laser beam is also recorded in order to subtract background profiles without the laser beam. Finally we adjust a power-law function $\propto r^{-s}$ to $T(r)$ in a region of large $r$ in order to determine the L\'evy exponent $s=3+\alpha$ \cite{Baudouin2014}.



\begin{figure}[h]
\centering

\includegraphics[scale=1.]{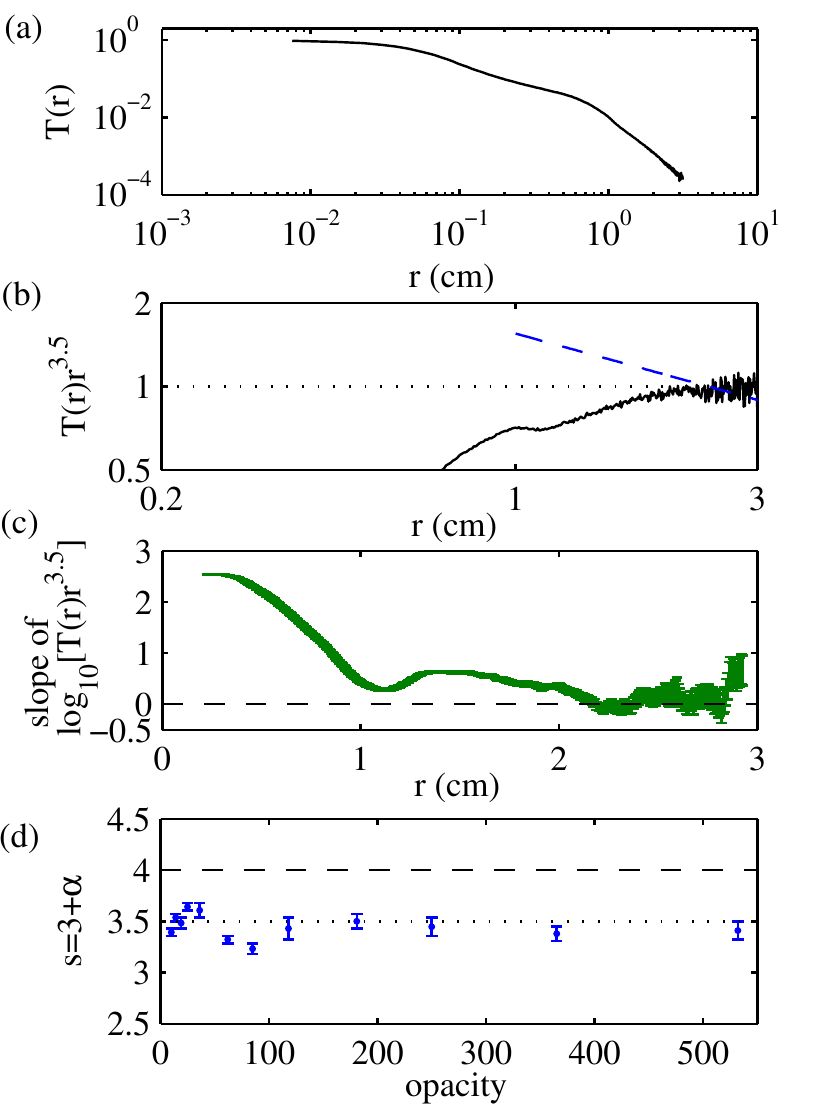}
\caption{
a) Experimental radial transmission profile $T(r)$ for an opacity of $O=360$. 
b) Rescaled experimental radial transmission  $T(r)r^{3.5}$ for an opacity of $O=360$ together with expected asymptotic results for CFR with a Doppler (blue dashed line) and Lorentz (black dotted line) profiles.
c) Local slope of the rescaled experimental transmission profile.
d) L\'evy exponent $s=3+\alpha$ (blue circles) obtained from a power law fit of the experimental radial profile $T(r)$ in the range $r\in[2.0,3.0]$ cm. The obtained values are $s=3.45\pm 0.13$. The black dotted and dashed lines are the theoretical values of $s=3.5$ for Lorentz profile and $s=4.0$ for Doppler profile, respectively.
}
\label{figure4}
\end{figure}

In Fig. \ref{figure4}a we plot the experimental radial transmission profile $T(r)$ in log-log scale obtained from the angle averaged distribution of a single image of the CCD camera. We have cut the radial profile around $r=3$ cm as for larger values of $r$ the $T(r)$ becomes smaller than background level. This also allows to avoid unwanted cell border effects. The observed radial profiles varies very little from image to image and have the same general behavior for the range of opacities explored. In Fig. \ref{figure4}b we also plot the rescaled experimental radial profile $T(r)r^{3.5}$ together with Lorentz ($T(r)r^{3.5}=1$) and Doppler ($T(r)r^{3.5}=r^{-0.5}$) asymptotic behavior. We clearly see that the experimental curve follows a Lorentz vapor behavior and is inconsistent with a Doppler vapor. In Fig. \ref{figure4}c we show a local slope of the rescaled profile $T(r)r^{3.5}$ and extract the asymptotic slope $s$ shown in Fig. \ref{figure4}d. The average value is $s=3+\alpha=3.45\pm 0.13$ (Fig. \ref{figure4}d), consistent with the expected  $\alpha=0.5$ value for a Lorentz vapor. This value is constant over the full range of opacity from $O=11$ to $O=530$ explored here (Fig. \ref{figure4}d). 

The geometry of the cell used in our experiment allows probing a range of radial distance with $r>3L$ such that the consideration leading to eq. \ref{radialTrans}, which consists in taking $1+\frac{L^2}{r^2}\approx1$, gives a good approximation. The detected photons are scattered inside the atomic vapor and thus have a restricted range of step lengths, corresponding to a truncation of the L\'evy flights being sensitive to non-diverging step length minimizing possible effects of truncation related to the cell thickness. Nevertheless, for the range of opacities explored in this work the relevant step lengths are larger than $10\bar{x}_0$, with $\bar{x}_0=1/n\sigma(0)$  the absorption penetration depth at line center, falling in the asymptotic limit $p(x)\propto x^{-1-\alpha}$, with $\alpha=0.5$ (see Fig. \ref{figure2}a and Supplemental Material \cite{Supp} for detailed discussion). This allows the measure of $T(r)$ to be an adequate technique to probe the L\'evy exponent.

\begin{figure}[h]
\centering
\includegraphics[scale=1.]{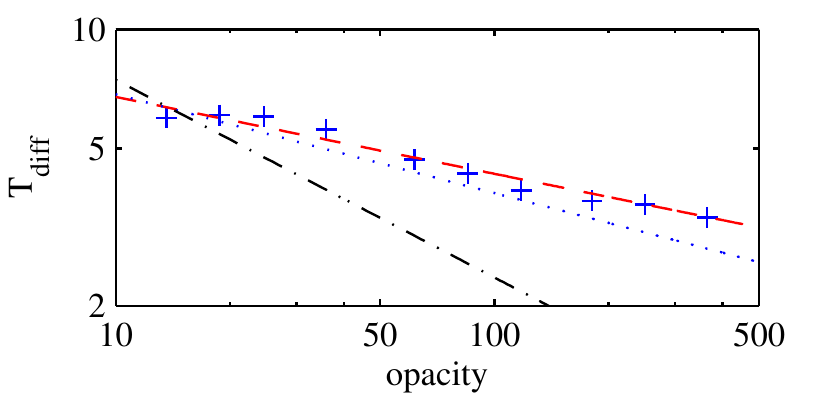}
\caption{
Normalized total transmission $T_\mathrm{diff}$ as a function of the opacity $O$. The blue data are obtained from the integral of the  experimental $T(r)$ (see the text). The red dashed line is a theoretical adjust with Eq. \ref{T_diff}.
The blue points correspond to a  L\'evy exponent of $\alpha=0.5$ and the dashed-point black curve would correspond to the total transmission in the Doppler broadened case.}
\label{figure5}
\end{figure}

For an alternative evaluation of the L\'evy exponent $\alpha$ via the scaling law given by Eq. \ref{T_diff}  \cite{Groth2012,Baudouin2014}, we also compute from these radial profiles the total diffusive transmission $T_{\mathrm{diff}}$ by integrating $T(r)$ over all $r$ up the the cell radius: $T_{\mathrm{diff}}=\int_0^{\infty}T(r)rdr$.  Experimental values of $T_\mathrm{diff}$ {\it vs} opacity $(O)$ are shown in Fig. \ref{figure5}, in log-log scale. By fitting the experimental values with Eq. \ref{T_diff}, we obtain for the exponent $\beta=\alpha/2=0.20\pm 0.03$, corresponding to a L\'evy exponent $\alpha=0.40\pm0.06$, in good agreement with the expected value for a Lorentz vapor.

The results for both radial and for total transmission are consistent with expected random walk for a Lorentz emission and Lorentz absorption. Using Monte-Carlo simulations we have checked that others combinations of absorption and emission profiles are not consistent with the results discussed here. For instance, for a Lorentz absorption and a Doppler emission, the result is a step length distribution that decays fast for large steps implying normal diffusion, as the Doppler emission profile decays much faster than the Lorentz absorption.


In summary, we have designed a experimental platform allowing to control the L\'evy exponent from $\alpha\approx 1$ with a diverging diffusion coefficient \cite {Mercadier2009,Mercadier2013,Baudouin2014} to $\alpha\approx0.5$ where even the average step size is infinite. This platform allows to study fundamental aspects of L\'evy flights and is of interest in a range of light scattering systems, including atomic clocks based on hot vapors as used in satellite navigation systems or for refined models in radiative transfer in astrophysics \cite{Hummer1968}. So far, we have operated these experiments in a steady state regime with low power in the linear optics limit. It will be interesting to extend such experiments into time dependent regimes, where the distribution of waiting time can lead to sub- and superdiffusive spreading in so-called L\'evy walks. Extending the present experiments into nonlinear optics do not pose important technical problems. If non classical light sources are used, the atomic systems might allow to study quantum correlations in L\'evy flights. 

\begin{acknowledgments}
We acknowledge financial support from the Brazilian Coordena\c c\~ao de Aperfei\c coamento de Pessoal de N\'ivel Superior (CAPES) and Conselho Nacional de Desenvolvimento Científico e Tecnológico (CNPq). T.P.S thanks financial support from Pronex/Fapesq-PB/CNPq. This  work was conducted within the framework of the project OPTIMAL granted by the European Union by means ofthe Fond Europ\'een de d\'eveloppement r\'egional, FEDER.
\end{acknowledgments}



\bibliography{Levy} 

\end{document}